\documentclass[12pt]{article} 

\usepackage{graphicx}
\usepackage{amssymb}
\usepackage{url}

\begin{document}

\title{Scanning of Vehicles for Nuclear Materials}


\author{J. I. Katz \\
Dept. Physics and McDonnell Center for the Space Sciences \\
Washington University, St. Louis, Mo. 63130}

\maketitle

\begin{abstract}
Might a nuclear-armed terrorist group or state use ordinary commerce to
deliver a nuclear weapon by smuggling it in a cargo container or vehicle?
This delivery method would be the only one available to a sub-state actor,
and it might enable a state to make an unattributed attack.  Detection of a
weapon or fissile material smuggled in this manner is difficult because of
the large volume and mass available for shielding.  Here I review methods
for screening cargo containers to detect the possible presence of nuclear
threats.  Because of the large volume of innocent international commerce,
and the cost and disruption of secondary screening by opening and 
inspection, it is essential that the method be rapid and have a low
false-positive rate.  Shielding can prevent the detection of neutrons
emitted spontaneously or by induced fission.  The two promising methods are
muon tomography and high energy X-radiography.  If they do not detect a
shielded threat object they can detect the shield itself.
\end{abstract}


\section{Introduction}
There are many means of delivering a nuclear device to a target.  Advanced
militaries may use aircraft, artillery, cruise missiles or ballistic
missiles (either kind of missile may be launched from land, sea, submarine
or air platforms).  Less sophisticated foes may use less sophisticated
means, such as civilian land, sea or air vehicles.  These have the
advantage that they are not likely to be recognized as threats and
therefore possess the ultimate stealth of hiding in plain sight.  Even a
state may choose to use a civilian delivery system in order to avoid
attribution---it is comparatively easy to track, and determine the origin
of, a military system, but not of a civilian object whose external
characteristics are those of the many millions of vehicles or containers of
innocent commerce.

We first distinguish between radiological dispersal devices and nuclear
explosives.  The latter are called nuclear weapons in common parlance,
though they are better defined as supercritical systems in order to include
nuclear explosives nominally intended for civilian engineering purposes,
unweaponized prototypes, developmental or experimental explosives and
improvised nuclear devices (IND) that might be assembled from diverted
fissile material by groups with limited resources.  These two categories
have almost nothing in common.

The enormous number of vehicles and intermodal cargo containers in innocent
commerce (more than 10,000,000 cargo containers enter the U.~S. each year)
means that inspection for nuclear materials must be fast and not introduce
significant delays in the flow of commerce.  Inspection must not interfere
with the smooth loading and unloading of ships, or the passage of trains or
trucks at ports of lading or entry.  The false alarm rate must also be low,
preferably $< 0.1\%$ and certainly $< 1\%$, because sending even a few
containers to secondary inspection (opening the container) is disruptive:
the container must be removed from the flow of traffic, taken where it can
be opened and entered, unloaded and inspected, and then, if innocent (as
almost all will be) returned to the logistic system.

Most intermodal containers have interior dimensions of 2.4 m$\times$2.4
m$\times$12 m (40$^\prime$) or 2.4 m$\times$2.4 m$\times$6 m (20$^\prime$)
and a nominal load limit (for either size) of 27 metric tons.  Heavy trucks
and rail cars have similar parameters.  This offers ample room and mass for
shielding.  The central problem of vehicle scanning is detecting threat
material that is likely to be heavily shielded.
\section{Radiological Dispersal Devices (RDD)}
These may be defined as devices to disperse highly radioactive material to
maximize human exposure and contamination of property \cite{ZL04}.  The
threat lies in the wide distribution of very strong radiation sources, often
only weakly secured.  A famous example was the system of unattended
navigational beacons on the Arctic shore of Russia, powered by radioisotope
thermal generators (RTG) containing 40,000 Ci of $^{90}$Sr (these have been
recovered and are no longer a potential threat).

Strong radiation sources include food and blood irradiators, RTG, sources
used in non-destructive evaluation (NDE; essentially, ``X-ray machines''
using nuclear $\gamma$-rays) and used reactor fuel awaiting permanent
disposal or reprocessing.  Manufacture of weaker sources in large number
(such as americium smoke detectors, polonium static neutralizers and 
radiopharmaceuticals) requires a large radioisotope inventory.  In some
applications, such as irradiators, radioisotopes may be, and are being,
replaced by electron beams, but no substitutes are feasible in other
applications.

The most important radioisotopes are $^{60}$Co, $^{90}$Sr, $^{137}$Cs,
$^{192}$Ir, $^{210}$Po, $^{238}$Pu and $^{241}$Am.  $^{60}$Co, $^{137}$Cs
and $^{192}$Ir are strong emitters of penetrating gamma rays.  As a result
they require heavy shielding for safe handling (a lesser issue for a suicide
bomber, but even he will require that he not be disabled before reaching his
target).  Sufficient radiation to be detected by a sensitive portal monitor
may penetrate even heavy shielding.  $^{90}$Sr and its short-lived daughter
$^{90}$Y are $\beta$-emitters, but also emit a significant $\gamma$-ray flux
by internal bremsstrahlung \cite{SAD71}.  In contrast, $^{210}$Po,
$^{238}$Pu and $^{241}$Am are $\alpha$-particle emitters with very weak or
nearly undetectable $\gamma$-ray activity, are easily shielded from portal
monitors, and pose little hazard unless in contact with the body.  They are
extremely toxic if ingested, with an acute fatal dose of roughly 10 mCi,
corresponding to a few $\mu$g of short-lived $^{210}$Po.  Inhaled particles
of $\alpha$-emitters may produce delayed lung cancer because the local
radiation dose at the particle may be large.

The question of shielding and detectability is complicated by the likelihood
that a terrorist will be sloppy, and spill or poorly shield the material he
obtains.  Hence, even if he has obtained an $\alpha$-particle emitter, it
may be readily detectable (and readily contaminate the environment, as was
observed after the poisoning of Litvinenko by $^{210}$Po).  In fact, 
$\alpha$-particle emitters are notorious for their ability to migrate
unless securely sealed, apparently as a result of scattering by emitted
$\alpha$-particles or the recoiling daughter isotopes (the daughters of the
principal $\alpha$-emitters are stable or only weakly radioactive
themselves).

Radioactive material might be dispersed either explosively or
non-explosively.  The term ``dirty bomb'' describes the use of explosives to
disperse radioactive material.  Non-explosive dispersal would not be a
bomb at all, but rather radioactive poisoning, the poison introduced into
the air, water, food supply, or some other widely distributed medium.  

It is difficult to estimate likely casualties, quite apart from the question
of how much material would be dispersed (are we concerned about 1 Ci, 100
Ci, or 10000 Ci, and of which isotope?).  Most estimates of casualties from
explosive dispersal find more fatalities from the explosive itself than from
the radioisotope.  People run away from an explosion, minimizing their
exposure.  Radiopoisoning is likely a greater threat to life because it
would not be detected until the first victims fell sick, generally some time
after ingestion or exposure, and were correctly diagnosed (radiopoisoning is
not what a physician would consider first when faced with a patient with
generalized malaise or an unusual combination of symptoms) or the
radioactivity itself were detected.  Most studies have concluded the larger
{\it social\/} impact would be abandonment of contaminated ground and
structures, especially if exacerbated by exaggerated public fear of
radiation and demands for ``maximum achievable'' decontamination without
regard to cost.  The subject was reviewed by Zimmerman and Loeb who describe
RDD as ``economic weapons'' \cite{ZL04}.

Searching for RDD is like looking for lost keys under a lamppost.  They
are intrinsically strong sources, handling by amateurs is likely to be
sloppy, and shielding is likely to be incomplete.  But they are not a
major threat to human life, and are only a secondary threat to property
and the functioning of a modern society.

\section{Fissile Material and Nuclear Explosives}
Nuclear explosives may range from highly engineered nuclear weapons (NW)
prepared by a state's weapons program to improvised nuclear devices (IND)
made by sub-state actors with limited resources, using stolen or diverted
material.  Any of these may be a threat; a sophisticated NW may be stolen
or diverted by a sub-state actor (``terrorist group''), or concealed or
disguised as peaceful cargo by a state to defeat or confuse attribution.

Despite the wide possible range of engineering sophistication, all these
have important features in common:
\begin{enumerate}
\item They must contain a minimum quantity of fissile material, $^{235}$U
or $^{239}$Pu (other fissile isotopes exist, but in much smaller quantity
and are unlikely to be involved).  
\item If successfully detonated, the explosive yield has a characteristic
value of order 10 kilotons, with the emission of about 0.5 mole/kiloton of
neutrons, production of a large quantity of highly radioactive fission
products, and activation of surrounding material by the neutrons.  The
fission product activity at a time $t$ after detonation may be estimated
at $10^7 (t/1\,{\rm day})^{-1.2}\,$Ci for $t < 6\,$months \cite{G62}.
An engineered device may be designed to produce significantly more or less
yield, and an IND may fizzle, depending on the skill of its designers and
builders, but the characteristic order of magnitude is determined by the
fundamental laws of physics and properties of the materials.
\item Fatalities are likely to be of order 100,000 if detonated in a dense
urban area, as was the case in Hiroshima and Nagasaki.
\item The explosion of a smuggled nuclear weapon or IND would be a surface
burst, producing intense activation and local fallout.  This is in contrast
to the Hiroshima and Nagasaki explosions that were airbursts at sufficient
altitude (600 m) that no surface material was swept up into the fireball,
minimizing activation and fallout.
\item The blast and thermal effects of a surface burst would be reduced, in
comparison to airbursts like those at Hiroshima and Nagasaki, because of
shielding (radiation and hydrodynamic) by topographic and cultural
(buildings) features.
\end{enumerate}
Nuclear explosions are the threat with which we must be concerned.
\subsection{Signatures}
The IAEA has defined ``Significant Quantities'' of fissile materials.  These
are the characteristic quantities (but not quantitative values) of fissile
materials necessary for nuclear explosives.  They are set by fundamental
physical properties.  The quantities found in actual explosives are
necessarily of this order of magnitude, but need not be equal or close to
these values.  They determine the signatures that must be detected when
vehicles or cargo containers are scanned.  The Significant Quantity of
weapons-grade $^{239}$Pu is defined as 8 kg, and that of highly enriched
uranium (HEU; 90\% $^{235}$U) is defined as 25 kg.

Signatures of fissile materials are of two classes.  They are radioactive,
and their emissions may, in principle, be detected.  But they also may be
detected by their other properties:
\begin{itemize}
\item High atomic number
\item High density:
\begin{itemize}
\item $\alpha$-Pu 19.86 g/cm$^3$
\item $\delta$-Pu 15.92 g/cm$^3$
\item U 19.1 g/cm$^3$
\end{itemize}
\item In metallic form for nuclear explosives they are likely to be found as
compact masses in order to minimize the escape of neutrons.
\end{itemize}

Fissile materials are not (in comparison to the isotopes that might be used
in a RDD) very radioactive.  The activities of significant quantities are:
\begin{itemize}
\item $^{239}$Pu
\begin{itemize}
\item 500 Ci of $\alpha$-particle activity.  $\alpha$-particles are
completely stopped (the attenuation is not exponential) by tens of microns
of matter, and effectively undetectable unless the plutonium is dispersed
into contact with a detector.
\item 13 mCi of $\gamma$-rays.  This is the strength of a typical laboratory
source.  The spectrum is complex, with several $\gamma$-ray energies, most
around 400 KeV, which are readily shielded by a few cm of lead.
\item Of order 30 $\mu$Ci of neutrons (the value depends on the fraction of
$^{240}$Pu and the geometry of the source, and is higher for diverted
products of power reactors and lower for high grade weapons material).
Neutrons are very penetrating, but the source is weak, and they may be
shielded by tens of cm of borated or lithiated plastic or wax.
\end{itemize}
\item HEU
\begin{itemize}
\item 50 mCi of $\alpha$-particle activity, which is insignificant.
\item 30 mCi of low energy (mostly 185 keV), $\gamma$-rays that are mostly
absorbed in the uranium itself and readily shielded.
\item 1 nCi of neutrons, which is insignificant.
\end{itemize}
\end{itemize}

The spontaneous radioactivity of fissile materials, even in threat
quantities, is low, except for their $\alpha$-activity, which is shielded
even by a sheet of paper or cm of air.  Only their neutron emission offers
some prospect of detection.
\section{Detection}
\subsection{Passive}
Passive detection is generally easy for RDD but difficult for fissile
material.  Here we consider the passive detection of plutonium.  The
penetrating radioactivity of HEU is so weak that its passive detection is
far beyond the realm of possibility.

Most (by volume or mass) cargo that enters the United States arrives in
intermodal containers, most of them by sea.  Containers crossing our land
borders, unless transshipped from outside the Americas, are not threats,
and transshipped containers from outside the Americas will also have have
undergone long sea voyages.  These voyages offer the opportunity to detect
weak sources of radioactivity by integrating signals over long times.

Because neutrons are the most penetrating radioactive emission, we consider
the detection of neutrons emitted by plutonium secreted in an intermodal
shipping container.  Fetter, {\it et al.\/} \cite{F90} defined a nominal
threat: 5 kg of plutonium emitting $4.5 \times 10^5$ n/sec, surrounded by
a 50 cm thick spherical moderating shell of nominal high explosive.

\begin{figure}
\begin{center}
\includegraphics[width=5in]{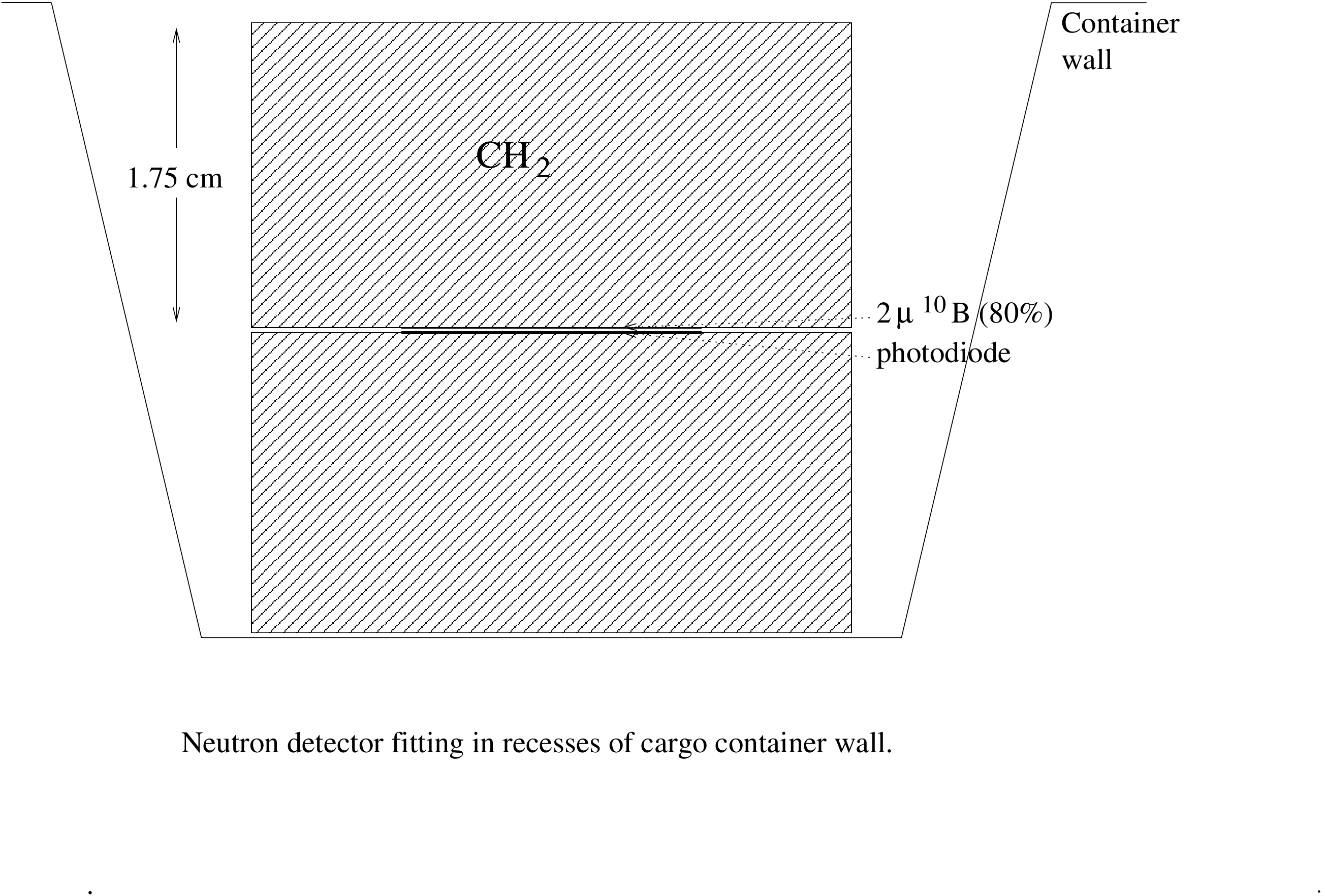}
\end{center}
\caption{\label{ndetect}}
\end{figure}

The detector consists of a 1 cm$^2$ silicon photodiode on which is deposited
2$\,\mu$ of 80\% enriched $^{10}$B boron.  This is sandwiched between two
1.75 cm thick slabs of paraffin moderator.  The entire package is just thin
enough to fit in the 3.75 cm deep recesses of the walls of a standard
intermodal container.  Neutrons escaping from the source are moderated by
the paraffin; at thermal energies they have a cross-section of nearly 4000 b
for the reaction $^{10}$B(n,$\alpha$)$^7$Li.  One of the charged particles
will be directed towards the photodiode and detected with an efficiency that
we conservatively take as 25\%.  The detector is shown in 
Fig.~\ref{ndetect}.  Such detectors, including processing circuitry on the
silicon substrate, can likely be mass-produced for less than \$10 each, and
powered by a 9 V battery.

We suppose three detectors on each 12$^\prime$ container, spaced so that no
point in the container is more than 3.5 m from a detector, and consider a
source at the center of the container and a detector half-way up the wall
but 3.29 m displaced along the length of the container, so that it is 3.5 m
from the source and the neutron flux falls obliquely on it.  We performed a
series of Monte Carlo calculations with the MCNP code to determine the
number of recorded counts in an integration time of $10^6$ s (about 12
days), corresponding to a transoceanic voyage with handling and local\
transportation at each end.  The results are shown in the Table.
\begin{table}
\begin{center}
\begin{tabular}{rlc}
No. & Features & Counts in $10^6$ s \\
1 & Baseline (4.22 cm Pu, 50 cm HE) & 340 \\
2 & Baseline + 26 innocent containers & 1600 \\
3 & Innocent container with cosmic rays & 0.6 \\
4 & 27 innocent containers with cosmic rays & 1.7 \\
5 & Baseline with 50 cm borated CH$_2$ shield & 0.08 \\
6 & Baseline + shield + 26 innocent containers & 0.7 \\
7 & No. 3 + shield & 0.6 \\
8 & No. 4 + shield & 1.8 \\
\end{tabular}
\end{center}
\caption{Results of Monte Carlo calculations \cite{K06} of neutrons detected
from a nominal source plutonium \cite{F90} in a cargo container under
conditions described in the text.}
\end{table}

The baseline case 1 consists of the nominal source 3.5 m from the described
1 cm$^2$ photodiode detector with paraffin moderator and $^{10}$B absorber.
Innocent containers, homogeneously filled with 30 tons of FeH, are in a
3$\times$3$\times$3 cuboidal array, with the source container at the center.
FeH is an approximation to nominal commercial traffic.  The heavy nucleus
(that is likely to be some mixture of carbon, oxygen, and metals) has little
effect; the hydrogen moderates and reflects neutrons, increasing the
detected flux roughly five-fold in case 2.

The detection of neutron emission in cases 1 and 2 is highly significant
compared to the background of cosmic ray generated neutrons in cases 3 and
4.  The higher signal in case 4 compared to case 3 is an example of the
``ship effect'': interaction of cosmic rays with surrounding matter (ship
and cargo) increases the background.  However, anyone sophisticated enough
to build a bomb, even a primitive IND or a shipment of illicitly obtained
fissile material, understands shielding.  In cases 5 and 6 the source is 
surrounded by an additional 50 cm thick spherical shield of 5\% natural
boron in paraffin.  This shielding reduces the signal even below background.
That suggests that the presence of shielding might be detected by the
``hole'' it leaves in the cosmic ray produced background.  Unfortunately,
cases 7 and 8, in which there is no source, but only the empty shielding
shell and the background source, shows that the shielding does not produce
a detectable ``hole'', probably because the shielding reflects neutrons
produced outside it.  The conclusion is that it is not possible to detect
the neutrons produced by a comparatively strong but shielded plutonium
source.  HEU sources are several orders of magnitude weaker.

These considerations also apply to active interrogation of targets by
photofission or neutron induced fission: Any induced fission neutrons can be
absorbed by shielding, even if the probe penetrates to the fissile material.
\subsection{Natural Interrogation---Muons}
Cosmic rays interacting with the Earth's atmosphere produce a downward flux
of muons, mostly $\mu^+$, of about 1/cm$^2$-minute at sea level.  Their mean
energy is about 3 GeV, with a Lorentz factor of about 30.  Muons have a
half-life in their rest frame of $2.2 \times 10^{-6}$ s, or a path length in
the atmosphere of about 20 km, allowing for relativistic time dilation,
roughly twice the altitude at which most of them are produced.  They lose
energy at a rate of about 2 MeV-cm$^2$/g by ionizing the matter through
which they pass, corresponding to a mean stopping length of about
1500 g/cm$^2$, about 1.5 times the atmospheric column density.  As a result,
the more energetic muons penetrate to sea level, through structures, cargo
containers, ships, and any other plausible obstacle, including threatening
pieces of fissile material and shielding.  This natural radiation is a
penetrating probe, without any additional exposure or man-made sources.

Fortunately, although muons are very penetrating, they scatter as they pass
through matter as a result of Coulomb interaction with atomic nuclei.  The
r.m.s. scattering angle of a charged particle of speed $\beta \equiv v/c$
and momentum $p$ (in GeV/c) after traversing a column density $L$ of a
material of radiation length $X$ (a quantity that is also related to the
bremsstrahlung and pair production by a relativistic electron) is \cite{R52}
\begin{equation}
\theta_0 = {14 \over \beta p} \sqrt{L \over X}\ {\rm mrad},
\end{equation}
where
\begin{equation}
X = {716 A \over Z (Z+1) \ln{(287/\sqrt{Z}})}\ {\rm g/cm^2}.
\end{equation}
Because of its inverse quadratic dependence on $Z$, $X$ is much smaller
for high-$Z$ materials than for lower $Z$ materials, and because of their
high density the characteristic length $X/\rho$ is even more sensitive to
$Z$.  For example, for iron $X/\rho = 1.86\,$cm, while for uranium $X/\rho =
0.31\,$cm.  Bodies of high-$Z$ material thus produce a characteristic muon
signature of high angular scattering concentrated in a compact volume.

\begin{figure}
\begin{center}
\includegraphics[width=5in]{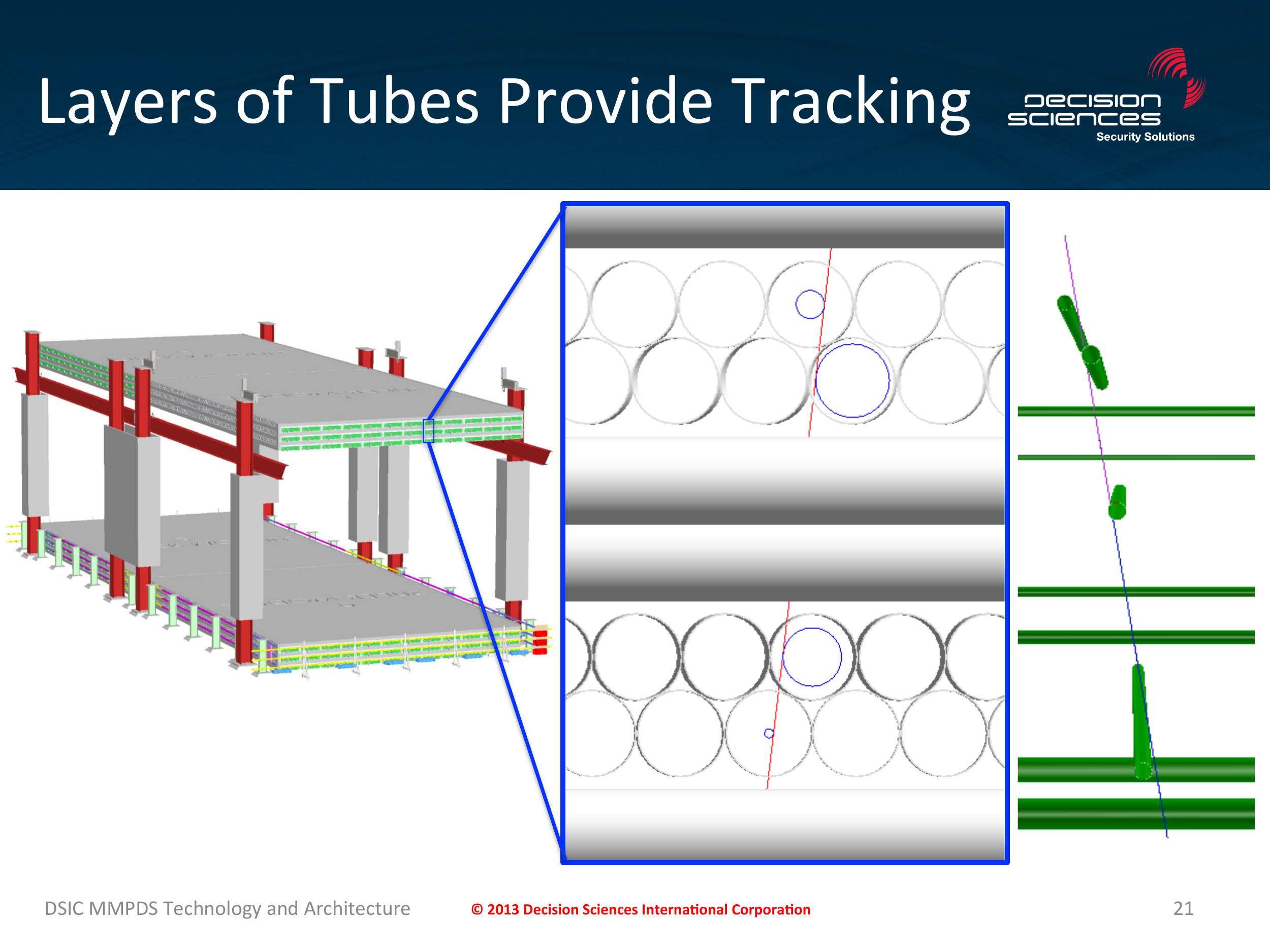}
\end{center}
\caption{Muon paths entering and leaving the sensed volume are determined
by timing information from orthogonal arrays of drift tubes.  Projection
to their points of closest approach indicates the scattering region.
(Decision Sciences International Corp.)}
\label{tubes}
\end{figure}

Tomographic reconstruction is necessary to localize this muon scattering.
The method was invented by Chris Morris \cite{M08}, and prototypes, up to
full scale for $40^\prime$ intermodal containers and tractor-trailer trucks,
have been developed by Decision Sciences International Corp.  The tracks of
muons are measured by orthogonal arrays of gas-filled drift tubes as they
enter and leave the vehicle or container under inspection, as shown in 
Fig.~\ref{tubes}.  Tracks are located to 0.25 mm accuracy by measuring the
time at which the ionization they create in the tubes is collected on the
anodes, wires running the lengths of the tubes, determining the distance
of closest approach of the track to the anode.  If the muon is only
scattered once in the instrumented volume the tracks will intersect, to the
accuracy of measurement, where it scattered.  If all the scattering is
concentrated in a small region of space, the incoming and outgoing tracks
will pass through that region.  This permits a three-dimensional tomographic
reconstruction of the distribution of scattering strength, which is an
indicator of the presence of dense bodies of high-$Z$ material.

\begin{figure}
\begin{center}
\includegraphics[width=5in]{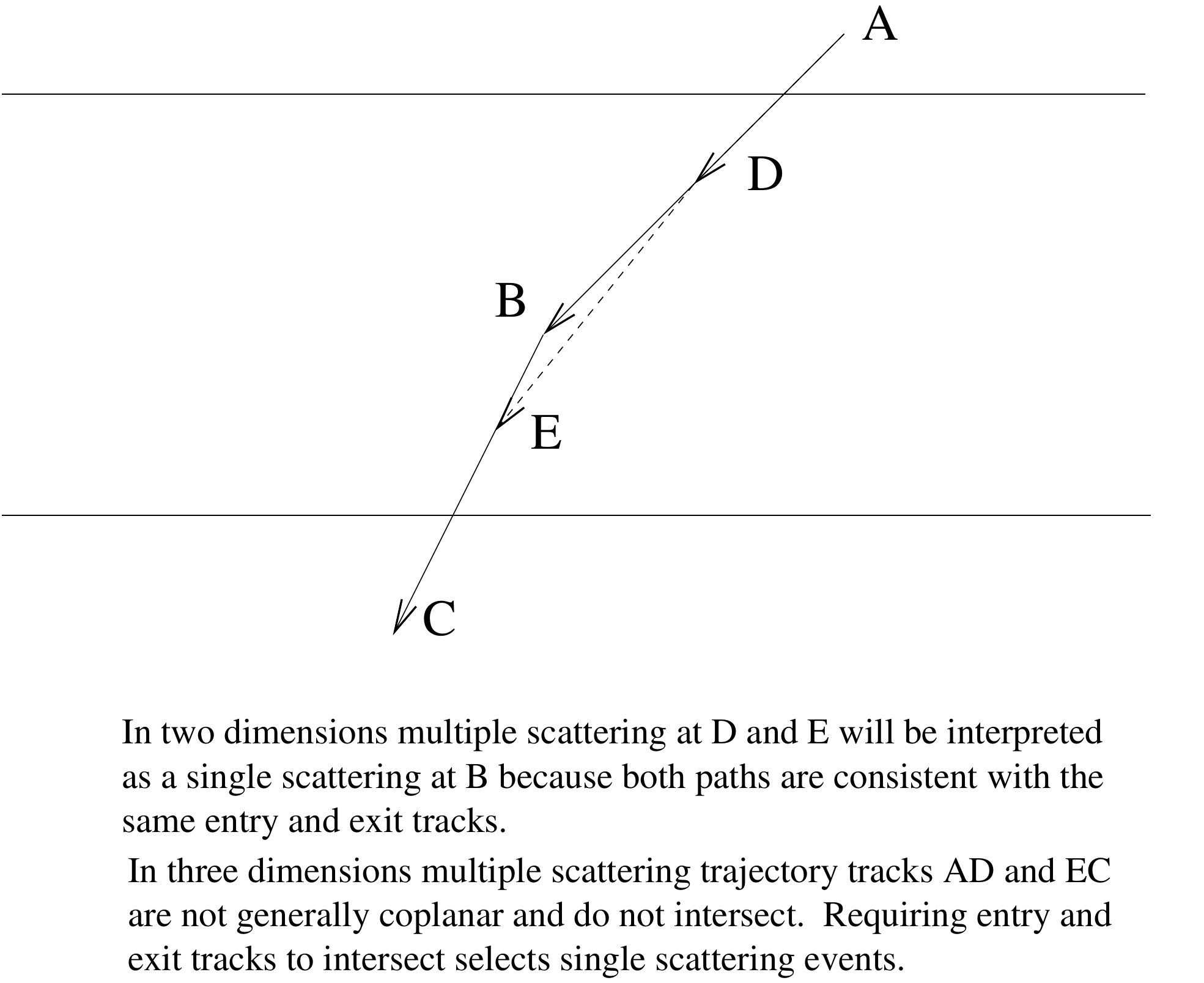}
\end{center}
\caption{Muon tomography works better in three than in two dimensions.} 
\label{2D3Dfig}
\end{figure}

Muon tomography is remarkable in that it works better in three dimensions
than in two.  In three dimensions it discriminates against multiple
scatterings that in two dimensions would be erroneously interpreted as a
single scattering at a location at which no actual scattering took place.
This is illustrated in Fig.~\ref{2D3Dfig}.  In two dimensions any two
tracks will intersect, and it is not possible to distinguish genuine
localized scattering at their intersection from a trajectory that scattered
in more than one place, neither at the intersection.  In three dimensions
the entry and exit tracks of a multiply scattered muon do not, in general,
intersect because they are generally not coplanar, and may be discriminated
by these means.

\begin{figure}
\begin{center}
\includegraphics[width=5in]{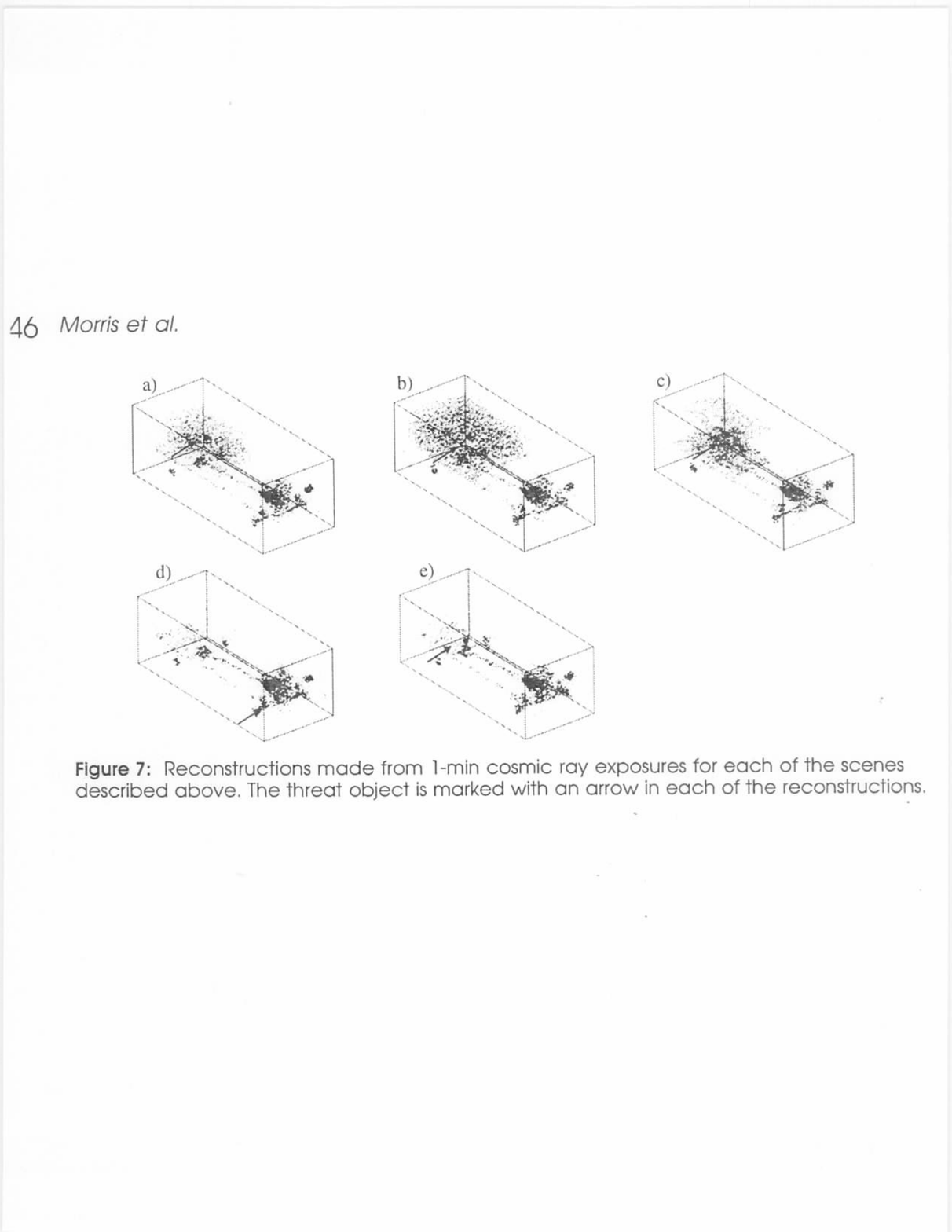}
\end{center}
\caption{Monte Carlo simulations of a cargo van with: a) A 3-foot stack of
$4^\prime \times 8^\prime$ sheets of plywood; b) 3.2 MT of miscellaneous
plastic, glass and steel clutter; c) A welding machine including two
horizontal $0.75^{\prime\prime}$ steel plates; d) Van without cargo but with
tungsten cube under engine block; e) Van without cargo but with tungsten
cube over differential.  \cite{M08}}
\label{cmvan}
\end{figure}

Morris, {\it el al.\/} \cite{M08} first demonstrated muon tomography in the
laboratory.  They simulated tomographic images of a 10-cm, 19 kg cube of
tungsten ($\rho = 19.25\,$g/cm$^{3}$, very similar to uranium, but with $Z =
74$ rather than 92) in a cargo van (Fig.~\ref{cmvan}).  The high-$Z$
material can be detected in one minute of integration, even when placed over
the differential and under the engine block, locations in which it is close
to large blocks of confusing iron.  Fig.~\ref{cmvan} also shows the power of
muon tomography to resolve and discriminate a variety of innocent cargoes.

\begin{figure}
\begin{center}
\includegraphics[width=5in]{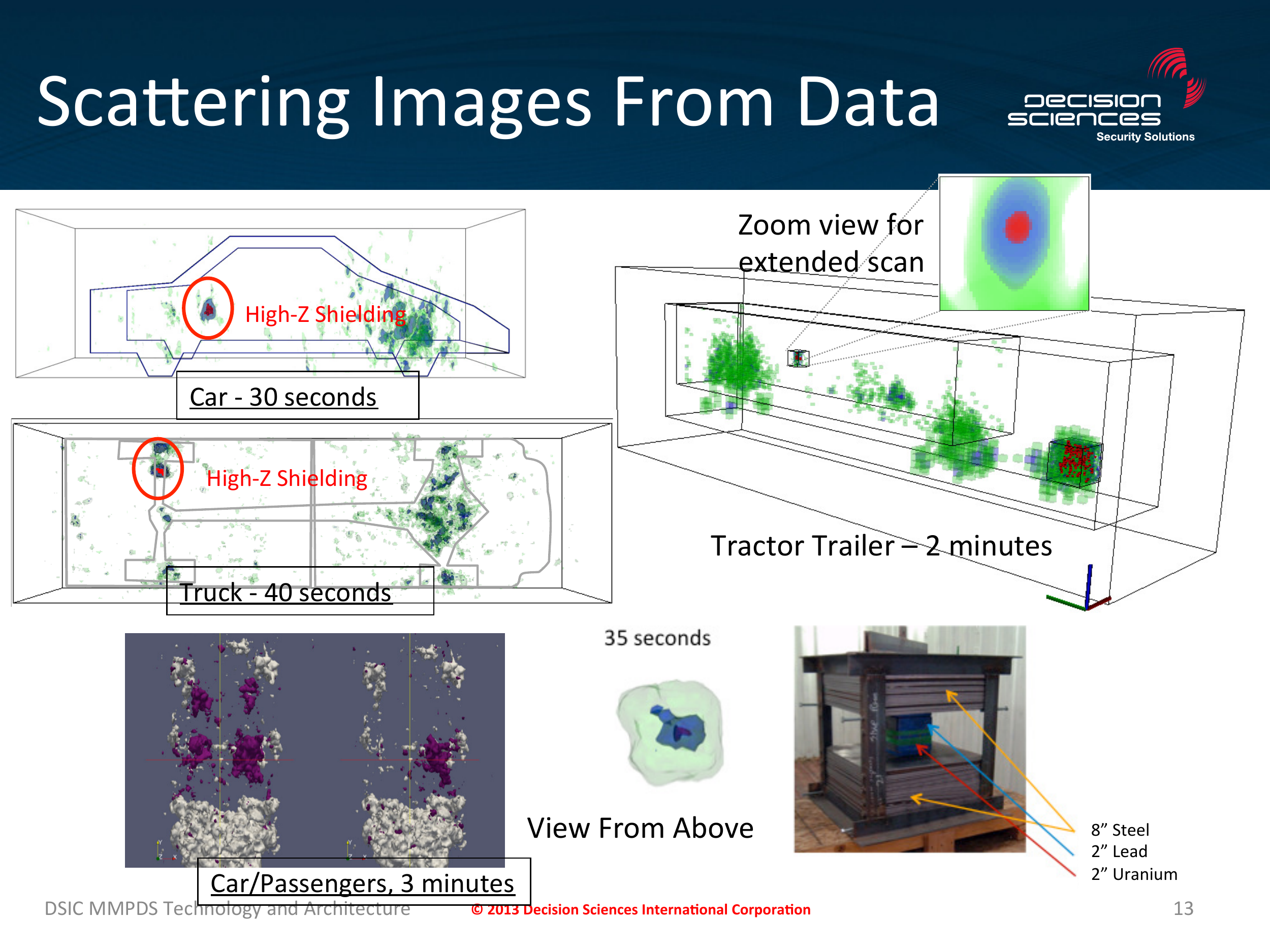}
\end{center}
\caption{Detection of mock threat objects by prototype muon tomography
systems.  In the car and truck the mock threat was 2 kg of uranium enclosed
in a hollow $5^{\prime\prime}$ cubic (18 kg) lead shield; in the
tractor-trailer it was 6 kg of uranium enclosed in a hollow
$7^{\prime\prime}$ cubic (58 kg) lead shield.  The drift tube arrays were
sized to the vehicles inside, and the larger system required to accommodate
the tractor-trailer has coarser resolution because the muon paths are 
longer (Decision Sciences International Corp.)}
\label{muprototype}
\end{figure}

Muon tomography has now been demonstrated on prototypes at scales up to
that large enough to accommodate an intermodal cargo container or
tractor-trailer.  Fig.~\ref{muprototype} shows the detection of mock threat
objects in an automobile, light truck and tractor-trailer.

The fundamental issue in muon tomography is a tradeoff between spatial
resolution and integration time because the muon flux is a natural
phenomenon and cannot be increased.  Higher spatial resolution is required
to detect smaller threat objects, increasing the required integration time.
Smaller threats also have shorter muon paths $L$, reducing the scattering
angle $\theta_0$.  Morris, {\it et al.\/} \cite{M08} present simulated
results, including ROC (false positive {\it vs.\/} false negative) curves
showing the tradeoffs.  

In mass screening operations, such as at a container port, the flow of
commerce sets an upper bound to the acceptable integration time.  Cranes
load and unload ships at a rate of approximately one container every 75
seconds.  If the integration time can be held to no more than about 60
seconds screening will not interfere with operations.  A container can be
rolled from the unloading crane, placed in the muon tomography chamber, and
rolled out in time for the next container; alternatively, this order can be
reversed in the port of embarkation.  Only the very few containers for which
a threat indication is found are removed for secondary inspection from the
flow path.  Operationally, minimizing the fraction of apparent positive
detections (probably to below 0.1\%) is critical; if this is not done, a
screening system will be considered unacceptable.

Because of the low radioactivity of fissile threats, they may not be 
surrounded by massive lead shields, unlike the mock threats in
Fig.~\ref{muprototype}.  Their $\gamma$-ray activity is low and easily
shielded with low- or medium-$Z$ material.  Neutron shields consist of
hydrogenous material with a small admixture of lithium or boron.  An IAEA
significant quantity (8 kg) of plutonium can be formed in a sphere 5 cm in
radius, and even the significant quantity of HEU (a 7 cm radius sphere) is
less than half the mass of the mock threat used in the tractor-trailer.

Bare fissile threat objects are harder to detect than massive lead shields
because they are smaller and scatter by smaller angles.  The resolution of
the tomographic system needs to be matched to the threat.  Scattering by an
under-resolved object will be spread over a larger voxel and the intense
scattering signature of dense high-$Z$ material will be lost.  On the other
hand, over-resolution requires a more expensive detector system with a more
drift tubes and longer integration times to detect a threat with acceptable
false positive and false negative rates.
\subsection{Active Interrogation---X-Rays}
\begin{figure}
\begin{center}
\includegraphics[width=5in]{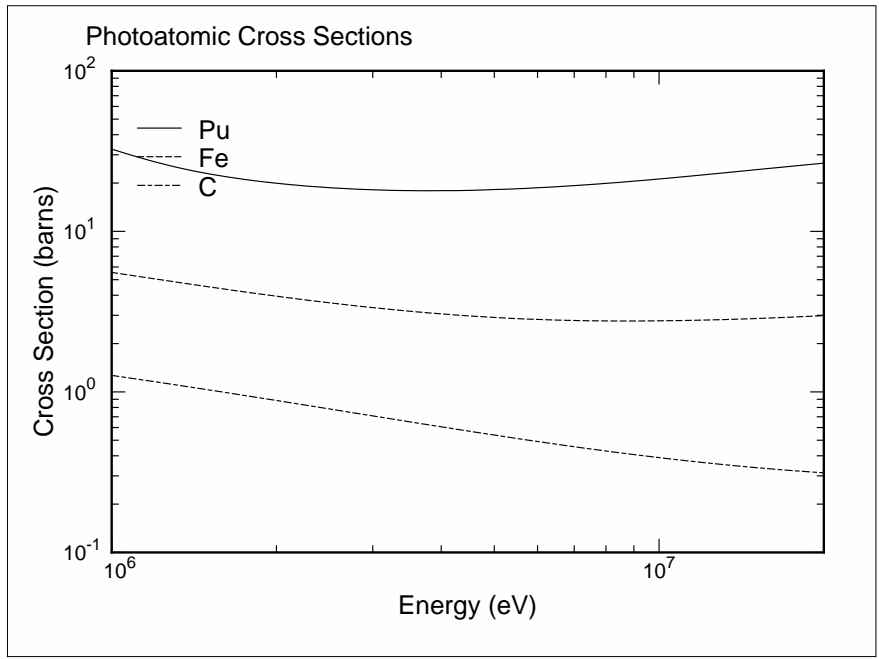}
\end{center}
\caption{Attenuation cross-sections per atom.  The cross-section at 5 MeV
for high-$Z$ material (uranium, plutonium) is an order of magnitude higher
than for medium-$Z$ (iron), the attenuation per unit length is about four
times as large and per gm/cm$^2$ about twice as large.  These ratios
increase with increasing X-ray energy because of the increase in pair
production by higher-$Z$ nuclei.  The cross-sections enter in an exponent
with a fairly large multiplier, so even small differences are important.
For example, 8 cm of $\delta$-plutonium attenuates 5 MeV X-rays by a factor
of 0.0029 but the same thickness of iron attenuates them by only a factor of
0.14, fifty times less. (t2.lanl.gov)} 
\label{xraysigma}
\end{figure}

A cargo container or vehicle may be actively probed with X-rays.  Compact
bodies of high-$Z$ material, such as masses of fissile material, have the
distinct X-radiographic signature of strong and spatially localized
attenuation.  Attenuation cross-sections as a function of energy are shown
in Fig.~\ref{xraysigma}.  

Because the cross-sections are large at low energy, radioactive $\gamma$-ray
sources (such at $^{60}$Co at 1.17 MeV and 1.33 MeV) that are used in
routine cargo screening are not sufficiently penetrating to distinguish
fissile threats from the enormous quantities of medium-$Z$ material in
innocent commerce.  The best discrimination between high-$Z$ and medium-$Z$
material is obtained by maximizing the X-ray energy.  This is done by using
high energy electrons from a linear accelerator to make a bremsstrahlung
X-ray source on a tungsten target.  The maximum usable electron energy (and
X-ray energy) is limited by the fact that X-rays of energies $\gtrsim
8\,$MeV photoproduce neutrons within the target (and elsewhere).  These
neutrons are emitted roughly isotropically and are difficult to shield; the
optimal electron energy is close to 10 MeV.  For representative parameters
at this energy the exposure of an unshielded operator at 20 m distance would
be about $1\,\mu$rem per container, or 100 mrem/year of full-time work at
one container per minute, less than natural backgrounds and 2\% of the
occupational limit \cite{KBBM07}.

The optimal irradiation geometry is an obliquely downward fan-beam produced
by a bremsstrahlung source above the vehicle or container to be scanned, 
detected by a line of collimated detectors (scintillators enclosed in
cylindrical holes oriented towards the X-ray source in a thick slab of
lead absorber) below the vehicle.  The vehicle or container would be pulled
through the radiographic system by a chain or conveyor (as in a car-wash),
without a driver inside.  At a representative electron accelerator pulse
rate of 100/s, a $40^\prime$ container can be scanned in 12 seconds with an
along-track resolution of 1 cm.  A linear array of 260 detectors, NaI(Tl)
scintillators measuring total deposited energy per pulse, provides 1 cm
cross-track resolution at readily achievable pulse strengths \cite{KBBM07}.

\begin{figure}
\begin{center}
\includegraphics[width=5in]{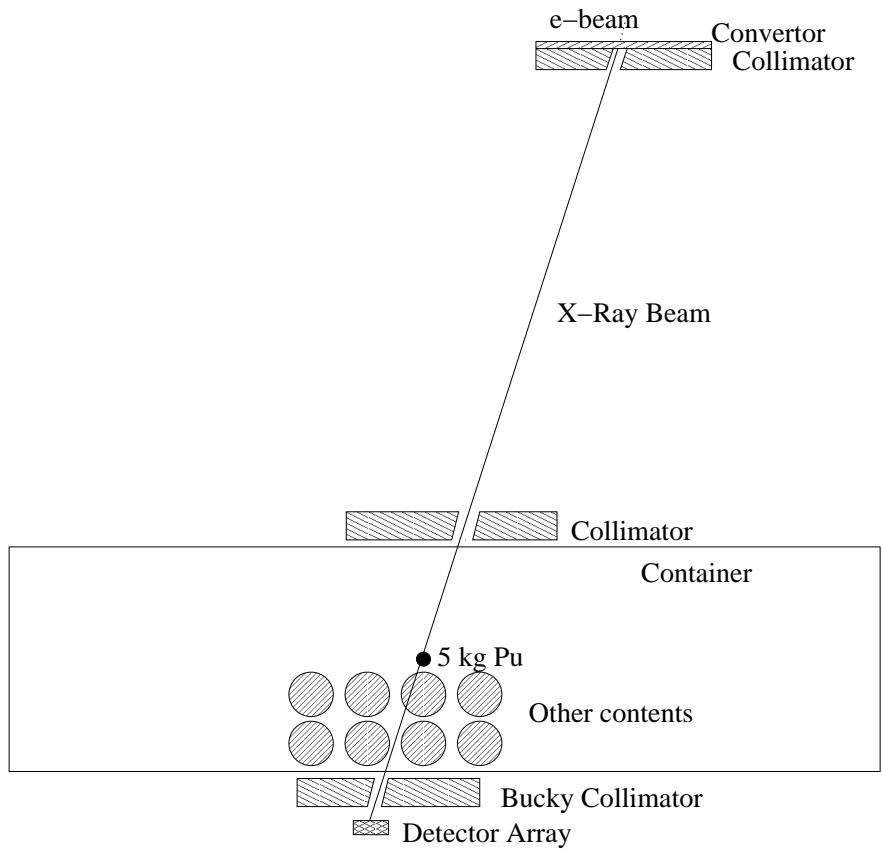}
\end{center}
\caption{Geometry of X-radiography of container.  The $40^\prime$ container
shown contains a 5 kg sphere of plutonium and 30 MT of half-density iron
spheres of 20 cm radius, representing automobile engine blocks or similar
medium-$Z$ clutter.}
\label{Xscan}
\end{figure}

This geometry, shown in Fig.~\ref{Xscan}, has several advantages.  Downward
illumination avoids false positive signals from end-on rod stock, long
ingots, railroad axles, shafts and similar long slender steel objects
because these will be laid horizontally on the floor.  The earth acts as a
beam-stop, minimizing the X-radiation dose to the surroundings (the operator
can be remote, if necessary).  Oblique illumination avoids false positives
from long slender vertical columns, and the use of two intersecting oblique
illumination directions permits localization of suspect objects andi
verification of their compactness in three dimensions \cite{Patent}.

\begin{figure}
\begin{center}
\includegraphics[width=5in]{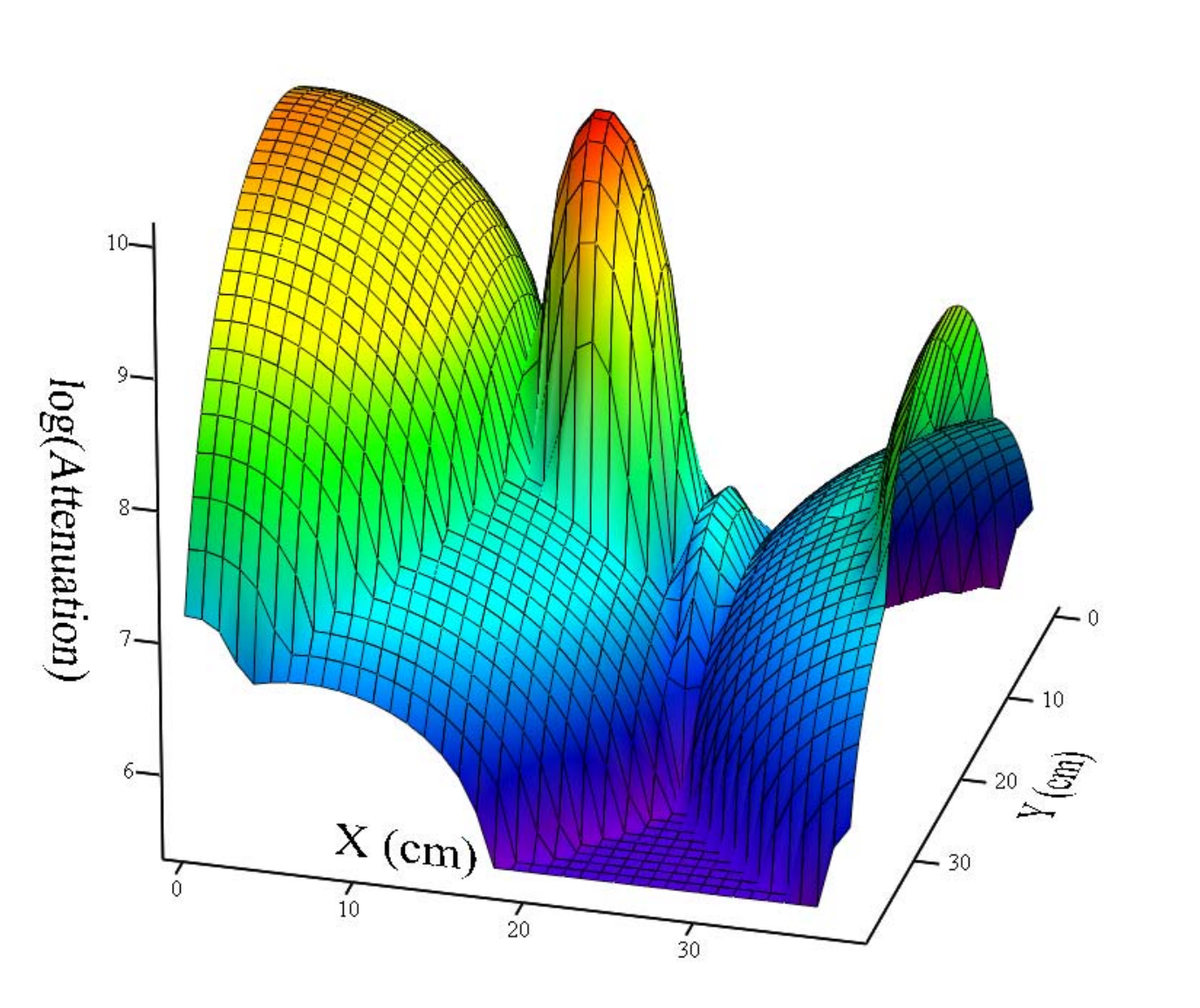}
\end{center}
\caption{X-radiograph of 5 kg plutonium sphere with clutter of 30 MT of
20 cm radius half-density iron spheres in $40^\prime$ containers.  The
compact peak of attenuation unambiguously indicates the presence of a
compact high-$Z$ object, a characteristic signature of a fissile threat.
X-irradiation is $13^\circ$ from vertical so the absorption maximum is
displaced horizontally from the sphere.  The zero of the attenuation scale
is arbitrary \cite{KBBM07}.}
\label{Xradimage}
\end{figure}

The results of X-radiography of the configuration of Fig.~\ref{Xscan}, as
computed by the MCNPX Monte Carlo code, are shown in Fig.~\ref{Xradimage}.
Details are given in \cite{KBBM07}.  The compact high-$Z$ fissile object (a
5 kg plutonium sphere) is apparent, and readily distinguished from the
clutter, even though the total attenuation through the clutter may exceed
that through the threat object.  Threat objects are distinguished by the
combination of their high attenuation and compact size.  

It is possible to surround fissile material with shielding so thick and
opaque that insufficient X-rays penetrate to reveal what is inside.  40 cm
of full-density iron (two diameters of the half-density spheres seen in
Figs.~\ref{Xscan} and \ref{Xradimage} are sufficient).  However, such
heavy shielding would be evident in X-radiography and would be a signal
that secondary inspection is necessary.  Such large compact single masses
of medium- or high-$Z$ material are rare in innocent commerce; when many
tons of metal are shipped, they are usually distributed as smaller bodies
across the floor of the container.





\section*{Acknowledgments}
I thank M. Sossong for discussions and providing the figures that depict
work at Decision Sciences International Corporation.

\end{document}